\documentclass[conference]{IEEEtran}
\IEEEoverridecommandlockouts
\usepackage{cite}
\usepackage{amsmath,amssymb,amsfonts}
\usepackage{graphicx}
\usepackage{textcomp}
\usepackage{xcolor}
\usepackage{algorithm}
\usepackage{algpseudocode}

\usepackage{booktabs,array}
\usepackage{subfigure}
\usepackage{bm}
\usepackage{comment}
\usepackage{makecell}
\usepackage{stfloats}
\def\BibTeX{{\rm B\kern-.05em{\sc i\kern-.025em b}\kern-.08em
    T\kern-.1667em\lower.7ex\hbox{E}\kern-.125emX}}
\allowdisplaybreaks[4]

\begin{document}

\title{RIS-Assisted Downlink Pinching-Antenna Systems With NLOS Environments  \\\vspace{-3mm}}
\author{\IEEEauthorblockN{Yue Geng, Kah Chan Teh, Tee Hiang Cheng}
	\IEEEauthorblockA{School of Electrical and Electronic Engineering, Nanyang Technological University \\  Email: yue.geng@ntu.edu.sg, ekcteh@ntu.edu.sg, ethcheng@ntu.edu.sg,  }\vspace{-4mm}
}
\maketitle

\begin{abstract}
This paper investigates a reconfigurable intelligent surface (RIS)-assisted downlink pinching-antenna system (PASS) in the presence of non-line-of-sight (NLOS) users caused by unavoidable obstacles. An RIS is deployed to extend the coverage of the physically constrained PASS by establishing virtual line-of-sight (LOS) links. To improve system performance, we propose a zero-forcing (ZF) beamforming-based alternating optimization (AO) framework that minimizes the transmit power by jointly optimizing the RIS reflection coefficients and pinching-antenna (PA) activation positions. Specifically, the RIS reflection coefficients are optimized using the Riemannian conjugate gradient (RCG) method, while a novel projected gradient-based method is developed to efficiently optimize the PA activation positions. Simulation results demonstrate that the proposed AO framework outperforms baseline schemes in terms of both transmit power and computational efficiency under different system configurations including the rate requirements and RIS sizes.
\end{abstract}


\section{Introduction}

To support the high data rates, massive connectivity, and wide-area coverage envisioned for future wireless networks, reconfigurable antenna technologies have recently attracted considerable attention, among which reconfigurable intelligent surfaces (RISs) and pinching-antenna systems (PASSs) are two promising technologies \cite{ra_sur,ris_sur,pa_sur}. An RIS enables programmable control of wireless propagation through a large number of passive reflecting elements with individually adjustable phase shifts. By properly configuring these elements, RISs can establish virtual line-of-sight (LOS) links, mitigate blockage, and create additional propagation paths with relatively low power consumption and hardware cost \cite{ris1}. In contrast, PASS introduces reconfigurability directly at the antenna side. A typical PASS consists of dielectric waveguides and multiple pinching antennas (PAs), whose positions can be flexibly adjusted along the waveguide \cite{pa1,pa2}. By adapting the PA positions to user locations and channel conditions, PASS provides additional spatial degrees of freedom (DoFs) compared with conventional fixed-position antenna systems and can significantly enhance LOS transmission \cite{pa3}. Moreover, the positions of multiple PAs can be jointly optimized to control their signal phases and realize pinching beamforming, further improving communication performance \cite{pa4,pa5}.

The technical relationship between RIS and PASS and their respective application scenarios has been discussed in existing studies \cite{pa_ris1}. Beyond exploiting these two technologies separately, integrating RIS with PASS offers a promising means of combining their complementary advantages. In \cite{pa_ris2}, an RIS was incorporated into a PASS-assisted communication system, where PASS enhanced the direct LOS channel while the RIS provided an additional reflected link with configurable phase shifts to further improve channel capacity of a single user. For a similar RIS-assisted PASS system, a graph neural network (GNN)-based method was proposed in \cite{pa_ris3} to jointly optimize the PA positions and RIS phase configuration.

However, existing studies on RIS-assisted PASS mainly consider scenarios in which users can be effectively served through LOS links. In practical deployments, PASS is typically mounted on physical structures, such as building walls, and its installation range is therefore inherently constrained. Consequently, some users may experience blocked direct links due to surrounding obstacles that cannot be avoided by changing PA activation position, resulting in non-line-of-sight (NLOS) transmission and substantial performance degradation. This limitation can significantly restrict the coverage capability of PASS. RIS provides a natural solution to this issue by creating controllable reflected links that can serve as virtual LOS paths around obstacles. Therefore, integrating RIS with PASS is particularly attractive in scenarios with NLOS users, as it can extend the effective service region of PASS beyond that supported by direct LOS transmission. Nevertheless, RIS-assisted PASS with NLOS environments and the corresponding system optimization remains largely unexplored.

Motivated by this gap, this paper investigates RIS-assisted downlink PASS systems in mixed LOS and NLOS environments. Unlike existing RIS-assisted PASS studies that mainly use the RIS to enhance available LOS transmission, we exploit the RIS to establish a virtual LOS link for the NLOS user, thereby extending the effective coverage of PASS. A key challenge in this setting is the optimization of PA activation positions, which simultaneously affect the LOS channels and the RIS-assisted channel to the NLOS user, and the PA positioning problem has not been studied in existing RIS-assisted PASS systems for mixed LOS and NLOS users. Accordingly, we formulate a power minimization problem for evaluating the system performance and develop an alternating optimization (AO) framework centered on PA positioning to enhance the performance gain achieved by the PASS.

\section{System Model and Problem Formulation}

\begin{figure}[htbp]
	\centering
	{\includegraphics[width=1\columnwidth]{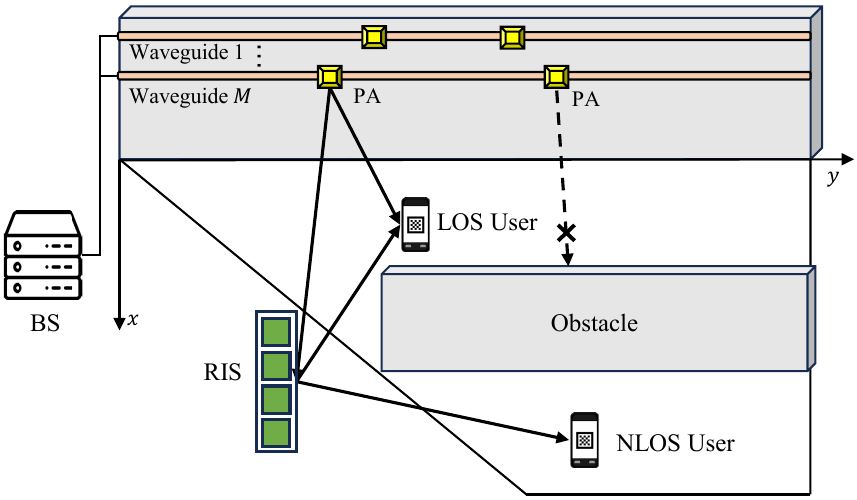}}
	\caption{The RIS-assisted downlink PASS with LOS and NLOS users.}
	\vspace{-4mm}
\end{figure}

As shown in Fig. 1, we consider an RIS-assisted downlink PASS. A base station (BS) transmits communication signals to multiple users through a PASS, which is deployed on an exterior wall parallel to the $y$-axis and consists of $M$ dielectric waveguides with length $A$. Each waveguide is equipped with $N$ PAs and is connected to the BS through an individual RF chain. The feed point and the $n$-th PA of the $m$-th waveguide are located at $\boldsymbol{\psi}_m^{0} = (0,0,h_m)$ and $\boldsymbol{\psi}_m^n = (0,\psi_m^n,h_m)$, respectively, where $\psi_m^n$ is controllable.  Denote the overall controllable PA position vector as $\boldsymbol{\psi}=\left[\psi_1^1,\dots,\psi_1^N,\dots,\psi_M^1,\dots,\psi_M^N\right]\in\mathbb{R}^{MN}$. 

Due to the presence of an unavoidable obstacle, the service area is divided into LOS and NLOS regions. We consider two single-antenna users located in the LOS and NLOS regions with positions denoted by $\mathbf{p}_{\rm L}=(x_{\rm L},y_{\rm L},0)$ and $\mathbf{p}_{\rm N}=(x_{\rm N},y_{\rm N},0)$, respectively. To extend the PASS coverage by establishing a virtual LOS link for the NLOS user, while simultaneously assisting the LOS user, a one-dimensional (1-D) RIS is deployed on the left side of the obstacle, parallel to the $x$-axis. The RIS consists of $L$ reflecting elements arranged as a uniform linear array (ULA) with half-wavelength spacing, and the position of its $l$-th reflecting element is denoted by $\mathbf{r}_l=(x_r+(l-1)\frac{\lambda}{2},y_r,z_r)$, where $(x_r,y_r,z_r)$ is the origin point of the RIS, and $\lambda$ is the signal wavelength.

Exploiting the spherical wave channel model, the channel between the LOS user and the PAs on the $m$-th waveguide is denoted by $\mathbf{g}^m_{\rm L}\in\mathbb{C}^N$, where the $n$-th coefficient is
\begin{align}
	\mathbf{g}^m_{\rm L}[n](\boldsymbol{\psi}_m^n,\mathbf{p}_{\rm L})=\frac{\eta^{\frac{1}{2}}e^{{-\jmath k_c}\lVert\boldsymbol{\psi}_m^n-\mathbf{p}_{\rm L} \rVert}}{\lVert\boldsymbol{\psi}_m^n-\mathbf{p}_{\rm L} \rVert},
\end{align}
where $\eta=\frac{\lambda^2}{16\pi^2}$ and $k_c=\frac{2\pi}{\lambda}$. Then, the overall channel between the PAs and the LOS user is
\begin{align}
	\mathbf{g}_{\rm L}=[(\mathbf{g}^1_{\rm L})^\top, \dots, (\mathbf{g}^M_{\rm L})^\top]^\top\in\mathbb{C}^{NM},
\end{align}
Similarly, the channel between PAs and $l$-th RIS element is given as 
\begin{align}
	\mathbf{g}_{{\rm R},l}=[(\mathbf{g}^1_{{\rm R},l})^\top, \dots, (\mathbf{g}^M_{{\rm R},l})^\top]^\top\in\mathbb{C}^{NM},
\end{align}
where
\begin{align}
	\mathbf{g}^m_{{\rm R},l}[n]=\frac{\eta^{\frac{1}{2}}e^{{-\jmath k_c}\lVert\boldsymbol{\psi}_m^n-\mathbf{r}_{l} \rVert}}{\lVert\boldsymbol{\psi}_m^n-\mathbf{r}_{l} \rVert},
\end{align}
and the overall PAs-RIS channel is 
\begin{align}
	\mathbf{G} = \left[\mathbf{g}_{{\rm R},1},\dots,\mathbf{g}_{{\rm R},L}\right]^\top \in \mathbb{C}^{L\times MN}.
\end{align}

We assume that the power allocated to each waveguide is equally distributed among all active PAs on it. Denote the in-waveguide channel between the BS and the PAs as
\begin{align}
	\mathbf{F}=\operatorname{Bdiag}(\mathbf{f}_1,\dots,\mathbf{f}_M)\in\mathbb{C}^{MN \times M},
\end{align}
where $\mathbf{f}_m\in\mathbb{C}^{N}$ is the in-waveguide channel between the feed point of the $m$-th waveguide and the PAs activated on it, and the $n$-th coefficient of which is obtained as
\begin{align}
	\mathbf{f}_m[n]=\frac{1}{\sqrt{N}}e^{-\jmath\frac{2\pi\lVert\boldsymbol{\psi}_m^n-\boldsymbol{\psi}_m^0\rVert}{\lambda_g}}=\frac{e^{-\jmath\frac{2\pi}{\lambda_g}\psi_m^n}}{\sqrt{N}},
\end{align}
where $\lambda_g=\frac{\lambda}{\eta_e}$ is the guided wavelength of the waveguide with effective refractive index $\eta_e$. In this paper, the in-waveguide propagation loss is omitted for simplicity as it is negligible compared with the free-space path loss \cite{pa2}.  

For the reflecting paths, the channels between the RIS and the LOS and NLOS users are respectively given as
\begin{align}
	\mathbf{h}_{\rm r,L}\!\! =\!\! \left[\!\frac{\eta^{\frac{1}{2}}e^{-\jmath k_c\left(\lVert\mathbf{r}_{1}-\mathbf{p}_{\rm L} \rVert\right)}}{\lVert\mathbf{r}_{1}-\mathbf{p}_{\rm L} \rVert},\!\dots\!,\frac{\eta^{\frac{1}{2}}e^{-\jmath k_c\left(\lVert\mathbf{r}_{L}-\mathbf{p}_{\rm L} \rVert\right)}}{\lVert\mathbf{r}_{L}-\mathbf{p}_{\rm L} \rVert}\!\right]^\top \!\!\!\!\!\in \!\mathbb{C}^L,
\end{align}

\begin{align}
	\mathbf{h}_{\rm r,N} \!\!=\!\! \left[\!\!\frac{\eta^{\frac{1}{2}}e^{-\jmath k_c\left(\lVert\mathbf{r}_{1}-\mathbf{p}_{\rm N} \rVert\right)}}{\lVert\mathbf{r}_{1}-\mathbf{p}_{\rm N} \rVert},\!\dots\!,\frac{\eta^{\frac{1}{2}}e^{-\jmath k_c\left(\lVert\mathbf{r}_{L}-\mathbf{p}_{\rm N} \rVert\right)}}{\lVert\mathbf{r}_{L}-\mathbf{p}_{\rm N} \rVert}\!\!\right]^\top \!\!\!\!\!\in\! \mathbb{C}^L.
\end{align} 

Denote the transmit data symbols to the LOS and NLOS users as $s_{\rm L}$ and $s_{\rm N}$, respectively, which are independent random variables with zero mean and unit variance. The signals are precoded by beamforming vectors $\mathbf{w}_k \in \mathbb{C}^{M}, \forall k \in\{ \rm L, N\}$. Then the signals received by the LOS and NLOS users are obtained as
\begin{align}
	y_{\rm L} = \left(\mathbf{h}_{{\rm r,L}}^\top\boldsymbol{\Phi}\mathbf{G}+\mathbf{g}_{\rm L}^\top\right)\mathbf{F}(\mathbf{w}_{\rm L}s_{\rm L}+\mathbf{w}_{\rm N}s_{\rm N})+ n_{\rm L},
\end{align}
and
\begin{align}
	y_{\rm N} = \mathbf{h}_{{\rm r,N}}^\top\boldsymbol{\Phi}\mathbf{G}\mathbf{F}(\mathbf{w}_{\rm L}s_{\rm L}+\mathbf{w}_{\rm N}s_{\rm N}) + n_{\rm N},
\end{align}
respectively, where $\boldsymbol{\Phi}=\operatorname{diag}(e^{\jmath\theta_1},\dots,e^{\jmath\theta_L})$ is the RIS reflection coefficient matrix with $\theta_l\in[0,2\pi]$, $n_k\sim\mathcal{CN}(0,\sigma^2), \forall k \in\{ \rm L, N\}$ is the received noise with power $\sigma^2$. Define the RIS reflection coefficient vector as $\boldsymbol{\phi}=[\phi_1,\dots,\phi_L]^\top=[e^{\jmath\theta_1},\dots,e^{\jmath\theta_L}]^\top\in\mathbb{C}^L$. Then the BS-LOS user and BS-NLOS user channels can be respectively denoted as $\mathbf{h}_{\rm L}^\top = \left(\boldsymbol{\phi}^\top\operatorname{diag}(\mathbf{h}_{{\rm r,L}})\mathbf{G}+\mathbf{g}_{\rm L}^\top\right)\mathbf{F}$ and $\mathbf{h}^\top_{\rm N} = \boldsymbol{\phi}^\top\operatorname{diag}(\mathbf{h}_{{\rm r,N}})\mathbf{G}\mathbf{F}$. The achievable rate of the LOS and NLOS users are then obtained as $R_{\rm L}=\log_2(1+\gamma_{{\rm L}})$ and $R_{\rm N}=\log_2(1+\gamma_{{\rm N}})$, where $\gamma_{{\rm L}}=\frac{\lvert \mathbf{h}^\top_{\rm L}\mathbf{w}_{\rm L}\rvert^2}{\lvert \mathbf{h}^\top_{\rm L}\mathbf{w}_{\rm N}\rvert^2+\sigma^2}$ and $\gamma_{{\rm N}}=\frac{\lvert \mathbf{h}^\top_{\rm N}\mathbf{w}_{\rm N}\rvert^2}{\lvert \mathbf{h}^\top_{\rm N}\mathbf{w}_{\rm L}\rvert^2+\sigma^2}$ are the signal-to-interference-plus-noise ratios (SINRs).

To accommodate the different quality-of-service (QoS) requirements of the LOS and NLOS users, their respective rate thresholds are denoted by $\Gamma_{\rm L}$ and $\Gamma_{\rm N}$, with the achievable rates required to satisfy $R_k \geq \Gamma_k, \forall k\in\{\rm L, \rm N\}$. When the RIS reflection coefficients and PA positions are fixed, obtaining the beamforming vectors for satisfying $R_k \geq \Gamma_k, \forall k\in\{\rm L, \rm N\}$ with optimized transmit power is a conventional transmit beamforming problem in multiuser multiple-input single-output (MU-MISO) systems, which can be solved using various approaches such as ZF beamforming, second-order cone programming (SOCP), or semidefinite programming (SDP) \cite{ris_sur,ris1,precode}. For simplicity, we adopt ZF beamforming in this work, while more advanced beamforming designs are left for future work. By defining $\mathbf{H}\!\! =\!\! [\mathbf{h}_{\rm L}\ \mathbf{h}_{{\rm N}}]^\top$ and $\beta_k=2^{\Gamma_k}-1,\forall k \in\{\rm L,N\}$, assuming that \(\mathbf H\) has full row rank, the ZF beamformer subject to $R_k \geq \Gamma_k, \forall k\in\{\rm L, \rm N\}$ with fixed $\boldsymbol{\phi}$ and $\boldsymbol{\psi}$ is
\begin{align}
	\label{zf}
	[\mathbf{w}_{\rm L}\ \mathbf{w}_{\rm N}] = \mathbf{H}^{\rm H}(\mathbf{H}\mathbf{H}^{\rm H})^{-1}\mathbf{D}_{\rm ZF},
\end{align}
where $\mathbf{D}_{\rm ZF} = \operatorname{diag}\left(\sqrt{\sigma^2 \beta_{{\rm L}}},\sqrt{\sigma^2\beta_{\rm N}}\right)$. Then the transmit power of the ZF beamformer is
\begin{align}
	\label{zfpower}
	P_{\rm ZF}(\boldsymbol{\phi},\boldsymbol{\psi}) = \operatorname{Tr}\left(\mathbf{D}_{\rm ZF}^2(\mathbf{H}\mathbf{H}^{\rm H})^{-1}\right),
\end{align}

To evaluate the system performance, we formulate the ZF transmit power minimization problem over the RIS reflection coefficients and PA activation positions. In particular, PA position optimization in the considered RIS-assisted mixed LOS and NLOS scenario has not been investigated in existing works, and is therefore the main focus of this work. The problem is formulated as 
\begin{subequations}
	\label{Q}
	\begin{IEEEeqnarray}{r,l}			
		$$\underset{\boldsymbol{\psi}, \boldsymbol{\phi}}{\min}$$&{\ P_{\rm ZF}(\boldsymbol{\phi},\boldsymbol{\psi}) } \\
		$$\operatorname{s.t.}$$
		&\ {\psi}_m^{n+1}-{\psi}_m^{n} \geq d_{\min}, \forall m, n \label{Qd2}\\
		&\ 0\leq{\psi}_m^n\leq A, \forall m,n \label{Crange2}\\
		&\ \lvert \phi_{l}\rvert=1, \ \forall l\in\{1,\dots,L\},    \label{Cris2}
	\end{IEEEeqnarray}
\end{subequations}
where \eqref{Qd2} enforces the minimum PA distance to avoid the coupling effect, \eqref{Crange2} confines the PAs on the waveguide, and \eqref{Cris2} denotes the unit-modulus constraint of the RIS reflection coefficients. The problem is highly non-convex due to the non-convex objective function, constraints, and the coupled variables. In the following section, we solve \eqref{Q} by developing the AO framework centered on PA position optimization to minimize the ZF transmit power.

\section{The Proposed AO-based Framework}
We develop the AO framework to solve \eqref{Q} in this section. Specifically, the RIS reflection coefficients and PA activation positions are alternatively optimized.

\subsection{RIS Reflection Coefficients Optimization}
When the PA positions are fixed as $\bar{\boldsymbol{\psi}}$, \eqref{Q} reduces to a conventional RIS reflection coefficient optimization problem, which has been extensively studied in the literature and can be addressed using various approaches, such as semidefinite relaxation (SDR), fractional programming (FP), or Riemannian optimization \cite{ris_sur,ris1,rcg1}. In this work, we adopt the Riemannian conjugate gradient (RCG) method to optimize the RIS reflection coefficients. With fixed $\boldsymbol{\psi}$, constraint \eqref{Cris2} can be interpreted as restricting the solutions to the Riemannian complex circle manifold (CCM), which is defined as
\begin{equation}
	\mathcal{M}_{\boldsymbol{\phi}}=\{\boldsymbol{\phi}\in \mathbb{C}^{L} \mid \lvert \boldsymbol{\phi}_{l}\rvert=1, \forall l = 1,\dots,L\},
\end{equation}
and the problem to be solved is
\begin{align}	
	\label{opt_phi}
	\underset{\boldsymbol{\phi}\in\mathcal{M}_{\boldsymbol{\phi}}}{\min} P_{\rm ZF} (\boldsymbol{\phi},\bar{\boldsymbol{\psi}}).
\end{align} 
Since $P_{\rm ZF}(\boldsymbol{\phi},\bar{\boldsymbol{\psi}})$ is a smooth function with respect to $\boldsymbol{\phi}$, \eqref{opt_phi} is an unconstrained Riemannian problem and can be solved via the standard RCG method. Since the RCG method for optimizing RIS reflection coefficients is well established in the literature, its implementation details are omitted for brevity, and interested readers are referred to \cite{rcg1,rcg2,HS}.

\subsection{PA Activation Position Optimization}
In this section, we optimize the PA activation positions with fixed $\bar{\boldsymbol{\phi}}$. The problem is formulated as 
\begin{subequations}
	\label{Q3}
	\begin{IEEEeqnarray}{r,l}			
		$$\underset{\boldsymbol{\psi}}{\min}$$&{\ P_{\rm ZF}(\bar{\boldsymbol{\phi}},\boldsymbol{\psi}) } \\
		$$\operatorname{s.t.}$$
		&\ {\psi}_m^{n+1}-{\psi}_m^{n} \geq d_{\min}, \forall m, n \label{Cd3}\\
		&\ 0\leq{\psi}_m^n\leq A, \forall m,n \label{Crange3},
	\end{IEEEeqnarray}
\end{subequations}
and we solve it by developing a projected gradient-based method in this section.

\subsubsection{Obtain the Gradient and Step Size}
Note that $P_{\rm ZF}(\bar{\boldsymbol{\phi}},\boldsymbol{\psi})$ is continuously differentiable with respect to $\boldsymbol{\psi}$, the corresponding gradient can be obtained as $\nabla_{\boldsymbol{\psi}}P_{\rm ZF}(\bar{\boldsymbol{\phi}},\boldsymbol{\psi}) = \partial P_{\rm ZF}(\bar{\boldsymbol{\phi}},\boldsymbol{\psi})/ \partial \boldsymbol{\psi}$ with
\begin{equation}
	\frac{\partial P_{\rm ZF}(\bar{\boldsymbol{\phi}},\boldsymbol{\psi})}{\partial\psi_m^n}
	=-2\sigma^2\Re\left\{
	\sum_{k\in\{\mathrm L,\mathrm N\}}
	\left[(\mathbf Q\mathbf D\mathbf Q\mathbf H)_{k,m}\right]^*
	\frac{\partial h_{k,m}}{\partial\psi_m^n}
	\right\},
\end{equation}
where $\mathbf Q=(\mathbf H\mathbf H^{\rm H})^{-1}$,
$\mathbf D=\operatorname{diag}(\beta_{\mathrm L},\beta_{\mathrm N})$, and $\frac{\partial h_{k,m}}{\partial\psi_{n,m}}, \forall k\in\{\mathrm L,\mathrm N\}$ are obtained based on the channel vector given in Section II. To update $\boldsymbol{\psi}^{t}$ at the $t$-th step, an unconstrained trial point is first obtained via a gradient-descent step as
\begin{align}
	\label{trial}
	\widetilde{\boldsymbol{\psi}}^t
	=
	\boldsymbol{\psi}^{t}
	-
	\alpha_t \nabla_{\boldsymbol{\psi}} P_{\rm ZF}
	\left(\bar{\boldsymbol{\phi}},\boldsymbol{\psi}^{t}\right),
\end{align}	
where $\alpha_t$ denotes the step size. We adopt the Barzilai–Borwein (BB) spectral step size, which exploits local curvature information from successive iterates to adaptively determine $\alpha_t$, enabling faster convergence than a fixed step size with negligible additional computational cost \cite{BB}. Define  $\mathbf{s}^{t-1}=\boldsymbol{\psi}^{t}-\boldsymbol{\psi}^{t-1}$ and $\mathbf{y}^{t-1}=\nabla P_{\rm ZF}(\bar{\boldsymbol{\phi}},\boldsymbol{\psi}^{t})	-\nabla_{\boldsymbol{\psi}} P_{\rm ZF}(\bar{\boldsymbol{\phi}},\boldsymbol{\psi}^{t-1})$, the BB step size can be obtained as
\begin{equation}
	\alpha_t^{\rm BB}
	=
	\left[\frac{(\mathbf{s}^{t-1})^{\top}\mathbf{s}^{t-1}}
	{(\mathbf{s}^{t-1})^{\top}\mathbf{y}^{t-1}}\right]_{\alpha_{\min}}^{\alpha_{\max}},
\end{equation}
which is safeguarded within a prescribed interval to avoid negative step size.

\subsubsection{Projection to the Feasible Set}
Note that $\widetilde{\boldsymbol{\psi}}^t$ may violate the PA feasibility constraints \eqref{Cd3}-\eqref{Crange3}, which can be projected onto the feasible set for constructing the search direction. The projection problem is formulated as
\begin{equation}
	\bar{\boldsymbol{\psi}}^t=\Pi_{\mathcal{C}}
	\left(
	\widetilde{\boldsymbol{\psi}}^t
	\right)
	=
	\arg\min_{\boldsymbol{\psi}\in\mathcal{C}}
	\frac{1}{2}\lVert
	\boldsymbol{\psi}
	-
	\widetilde{\boldsymbol{\psi}}^t
	\rVert^2,
\end{equation}
where $\mathcal{C}\!\triangleq\!\! \{\boldsymbol{\psi} \vert 0\leq{\psi}_m^n\leq A, {\psi}_m^{n+1}-{\psi}_m^{n} \geq d_{\min},\forall m,n\}$. Although this is a convex problem and can be solved using standard convex optimization methods, their computational complexity may be high for implementation in each update step. Note that the constraints for different waveguides are separable, the projection can be independently performed for each waveguide. For the $m$-th waveguide, define the variable $\boldsymbol{x}_m=[x_m^1,\dots,x_m^N]\in\mathbb{R}^{N}$ as
\begin{equation}
	x_m^n=\psi_m^n-(n-1)d_{\min}.
\end{equation}
Then, the constraints $\psi_{m}^{n+1}-\psi_m^{n}\ge d_{\min}$ and $0 \!\leq \psi^n_m\leq A$ for each waveguide are equivalently converted into the monotonicity constraint
\begin{equation}
	\label{mono1}
	\boldsymbol{x}_m\in\mathcal{U}\triangleq\{\boldsymbol{x}_m\in\mathbb{R}^{N}\vert0\!\le\! x_{m}^1\!\le\! x_{m}^2\!\le\!\cdots\!\le\! x_{m}^N \!\le\! B\},
\end{equation}
where $B=A-(N-1)d_{\min}$, and the projection problem for the $m$-th waveguide is converted as
\begin{equation}
	\label{proj}
	\bar{\boldsymbol{x}}_m^t
	=
	\Pi_{\mathcal{U}}(\boldsymbol{z}_m^t)
	=
	\arg\min_{\boldsymbol{x}_m\in\mathcal{U}}
	\frac{1}{2}
	\left\|
	\boldsymbol{x}_m-\boldsymbol{z}_m^t
	\right\|^2,
\end{equation}
where $\boldsymbol{z}_m^t=[z_{m,1}^t,\dots,z_{m,N}^t]^\top\in\mathbb{R}^N$ and $z^t_{m,n}=(\widetilde{\boldsymbol{\psi}}^t)_m^n - (n-1)d_{\min}$. Note that \eqref{proj} is a bounded isotonic regression problem, which can be efficiently solved using the pool-adjacent-violators algorithm (PAVA) with low computational complexity \cite{PAVA}. PAVA iteratively merges adjacent blocks that violate the monotonicity constraint and updates their values using clipped block means until a feasible projection is obtained. The detailed procedure is summarized in Algorithm \ref{alg:PAVA}, and the optimal $\bar{\boldsymbol{\psi}}^t$ can be obtained by applying Algorithm \ref{alg:PAVA} to each of the $M$ waveguides independently.

\begin{algorithm}[t]
	\caption{The PAVA Algorithm for Solving \eqref{proj}}
	\label{alg:PAVA}
	\begin{algorithmic}[1]
		\Require $\boldsymbol{z}_m^t$, $B$, and $d_{\min}$.
		\State Initialize the number of blocks as $q=0$.
		\For{$n=1,\ldots,N$}
		\State \parbox[t]{0.92\linewidth}{Set $q\leftarrow q+1$ and initialize a new block
			$\mathcal{B}_q=\{n\}$ with
			$v_q=z_{m,n}^t$ and $w_q=1$.}
		\While{$q>1$ and $v_{q-1}>v_q$}
		\State Merge the adjacent blocks as
		$\mathcal{B}_{q-1}\leftarrow
		\mathcal{B}_{q-1}\cup\mathcal{B}_q$.
		\State Update the block value as
		\[
		v_{q-1}\leftarrow
		\frac{w_{q-1}v_{q-1}+w_qv_q}
		{w_{q-1}+w_q}.
		\]
		\State Update
		$w_{q-1}\leftarrow w_{q-1}+w_q$
		and $q\leftarrow q-1$.
		\EndWhile
		\EndFor
		\For{$j=1,\ldots,q$}
		\State Clip the block value as
		$v_j\leftarrow [v_j]_{[0,B]}$.
		\State Set
		$\bar{x}_{m,n}^{t}=v_j$,
		$\forall n\in\mathcal{B}_j$.
		\EndFor
		\State Recover
		$\bar{\psi}_{m,n}^{t}
		=\bar{x}_{m,n}^{t}+(n-1)d_{\min}$,
		$\forall n$.
		\Ensure
		$\bar{\boldsymbol{\psi}}_m^t
		=[\bar{\psi}_{m,1}^{t},\ldots,
		\bar{\psi}_{m,N}^{t}]^\top$.
	\end{algorithmic}
\end{algorithm}

\subsubsection{Update the PA Positions}
The solution of the projection is obtained as $\bar{\boldsymbol{\psi}}^t=[(\bar{\boldsymbol{\psi}}_1^t)^\top,\dots,(\bar{\boldsymbol{\psi}}_M^t)^\top]^\top$. Accordingly, the update direction is obtained as 
\begin{align}
	\label{d}
	\mathbf{d}^t = \bar{\boldsymbol{\psi}}^t - \boldsymbol{\psi}^t.
\end{align}
By the variational inequality of Euclidean projection, \eqref{d} is a descent direction of $P_{\rm ZF}(\bar{\boldsymbol{\phi}},\boldsymbol{\psi})$ at $\boldsymbol{\psi}^t$. Moreover, since both $\bar{\boldsymbol{\psi}}^t$ and $\boldsymbol{\psi}^t$ belong to the convex feasible set $\mathcal{C}$, any update along this direction with a step size in $(0,1]$ remains feasible. We then employ Armijo backtracking line search to select a step size, with the sufficient decrease condition given as
\begin{align}
	\label{armijo}
	P_{\rm ZF}(\bar{\boldsymbol{\phi}},\boldsymbol{\psi}^t \!\!+\!\! \beta \mathbf{d}^t) \!\leq\! P_{\rm ZF}(\bar{\boldsymbol{\phi}},\boldsymbol{\psi}^t) \!+\! c\beta\nabla_{\boldsymbol{\psi}} P_{\rm ZF}^\top(\bar{\boldsymbol{\phi}},\boldsymbol{\psi}^t) \mathbf{d}^t.
\end{align}
The step size $\beta$ is determined by backtracking from an initial value of 1 until the Armijo condition is satisfied. The PA positions are then updated as $\boldsymbol{\psi}^{t+1} = \boldsymbol{\psi}^t + \beta \mathbf{d}^t$.

\subsection{Initialization and Summary of the AO Method}

A 1-D search method is utilized for the initialization of the AO method. Specifically, the PA positions are initially placed symmetrically around the RIS along each waveguide to minimize the path loss of the PA-RIS channel with a uniform spacing of $d_{\rm ini}\lambda$. Given these initial PA positions, the RIS reflection coefficients are first optimized using RCG method. Subsequently, a 1-D search is sequentially performed for each waveguide by uniformly sampling $S$ candidate PA-group centers between the RIS and the LOS user, while keeping the PA spacing and RIS reflection coefficients fixed. A candidate position is accepted only if it reduces the required ZF transmit power. The initialization aims to achieve a favorable balance between the LOS and NLOS users by accounting for their different channel conditions, while the initial PA spacing provides sufficient flexibility for subsequent PA position optimization.
\begin{algorithm}[t]
	\caption{The Proposed AO Method for Solving \eqref{Q}}
	\label{alg:AO}
	\begin{algorithmic}[1]
		\Require Initialize $\boldsymbol{\phi}^1$ and $\boldsymbol{\psi}^1$ using the procedure described in Section III-C.
		\For{$i=1:I$}
		\State Update $\boldsymbol{\phi}^{i+1}$ by solving \eqref{opt_phi} via RCG with fixed $\boldsymbol{\psi}^i$.
		\State Set $\boldsymbol{\psi}^{i,1}
		=\boldsymbol{\psi}^i$.
		\For{$t=1:T$}
		\State Obtain $\widetilde{\boldsymbol{\psi}}^{i,t}$
		via \eqref{trial} with the BB step size.
		\For{$m=1:M$}
		\State Obtain $\bar{\boldsymbol{\psi}}_m^{i,t}$
		via Algorithm \ref{alg:PAVA}.
		\EndFor
		\State Obtain the updated $\boldsymbol{\psi}^{i,t+1}$ via \eqref{armijo}.
		\If{$\lvert P_{\rm ZF}(\boldsymbol{\phi}^{i+1},\boldsymbol{\psi}^{i,t+1})-P_{\rm ZF}(\boldsymbol{\phi}^{i+1},\boldsymbol{\psi}^{i,t}) \rvert\!<\! \xi$}
		\State Break the inner loop.
		\EndIf
		\EndFor
		\State Update the PA positions as $\boldsymbol{\psi}^{i+1}\leftarrow\boldsymbol{\psi}^{i,t+1}$.
		\If{$\lvert P_{\rm ZF}(\boldsymbol{\phi}^i,\boldsymbol{\psi}^i)
			-
			P_{\rm ZF}(\boldsymbol{\phi}^{i+1},
			\boldsymbol{\psi}^{i+1})
			\rvert<\xi$}
		\State \textbf{break}.
		\EndIf
		\EndFor
		\Ensure Optimized $\boldsymbol{\phi}$ and
		$\boldsymbol{\psi}$.
	\end{algorithmic}
\end{algorithm}
The total AO method for solving \eqref{Q} is summarized as Algorithm \ref{alg:AO}. Since the Armijo line searches employed in both the RCG and the proposed projected gradient-based methods ensure a non-increasing objective value, and the transmit power is lower-bounded by zero, the proposed AO algorithm generates a monotonically non-increasing and convergent objective sequence. 

For the computational complexity, let $T_\phi$ and $T_\psi$ denote the numbers of RCG and projected gradient iterations, respectively, and let $B_\phi$ and $B_\psi$ denote the numbers of Armijo line-search steps. The ZF power and its gradients with respect to $\boldsymbol{\phi}$ and $\boldsymbol{\psi}$ can each be computed with complexity $\mathcal{O}(LMN)$. Consequently, the RCG optimization requires $\mathcal{O}(T_\phi(1+B_\phi)LMN)$ operations. For the projected gradient-based method, the BB step-size update and PAVA projection require $\mathcal{O}(MN)$ operations per trial, yielding an overall complexity of $\mathcal{O}(T_\psi(1+B_\psi)LMN)$. Therefore, with $I$ AO iterations, the overall complexity is $\mathcal{O}(I[T_\phi(1+B_\phi)+T_\psi(1+B_\psi)]LMN)$. Additionally, the initialization performs a 1-D search over $S$ candidates for each waveguide, requiring $\mathcal{O}(S LM^2N)$ operations.

\section{Simulation Results}

Simulation results are provided in this section to evaluate the performance of the RIS-assisted PASS systems and the proposed AO framework. Unless otherwise specified, the default simulation parameters are set as follows. The PASS has a length of $A=40$ and consists of $M=4$ waveguides, each supporting $N=4$ PAs. The waveguides are deployed parallel to the $y$-axis at heights $h_m=2+0.1(m-1)$, $m=1,\ldots,M$. The minimum PA distance is $d_{\min}=\frac{\lambda}{2}$. The LOS and NLOS users are randomly located within the rectangular regions defined by $x_{\rm L}\in[0,10]$, $y_{\rm L}\in[10,40]$ and $x_{\rm N}\in[10,20]$, $y_{\rm N}\in[20,40]$, respectively. The RIS reference point is located at $(x_r,y_r,z_r)=(10,10,2)$. The number of RIS reflecting elements is set to $L=36$. The carrier frequency is 28 GHz, and the refractive index of the waveguides is set to $\eta_e=1.4$. The receiver noise power at each user is set to $\sigma^2=-90$ dBm. For the proposed AO framework, the 1-D initialization search employs $d_{\rm ini}=4$ and $S=20$. The maximum numbers of outer AO iterations and inner iterations are set to $I=10$ and $T_\psi=T_\phi=100$, respectively. The maximum number of Armijo backtracking steps is set to $B_\psi=B_\phi=10$. The convergence threshold is $\xi=0.0001$. All simulation results are averaged over 1024 independent realizations of randomly generated LOS and NLOS user locations.

To evaluate the performance of the RIS-assisted PASS and the effectiveness of the proposed AO method, we implement the following baseline schemes for performance comparison. For the first three schemes, the PAs on each waveguide form a half-wavelength-spaced ULA, while the RIS phase shifts
are optimized via RCG.
\begin{itemize}
	\item ``\textbf{Near RIS}": The PA array is placed
	closest to the RIS to reduce the cascaded NLOS link path loss.
	\item ``\textbf{Near LU}": The PA array is placed
	closest to the LOS user to reduce its path loss.
	\item ``\textbf{Midpoint}": The PA array is centered
	at the midpoint between the $y$-coordinates of the
	LOS user and RIS to balance their path losses \cite{pa2}.
	\item ``\textbf{1-D Search}": Each PA position is
	optimized sequentially via element-wise 1-D search
	to minimize the ZF transmit power \cite{pa5}.
	The search runs for $10$ rounds with $500$ candidate
	positions per PA per round, and the RIS reflection coefficients
	are updated via RCG after each round.
\end{itemize}

\begin{figure}[t]
	\centering{\includegraphics[width=0.8\columnwidth]{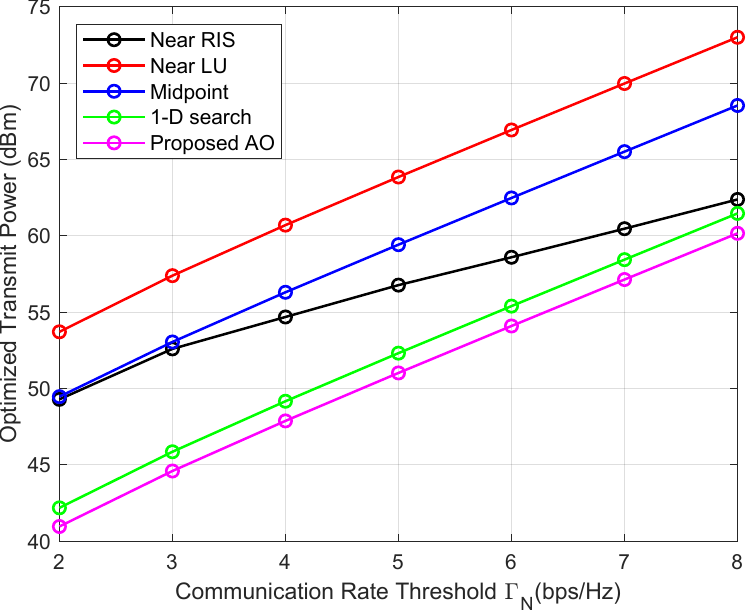}}
	\caption{The optimized transmit power versus the communication rate threshold of the NLOS user with $\Gamma_{\rm L}=8$ bps/Hz.}
	\vspace{-1.5mm}
	\label{fig:gamma}
\end{figure}

Figure~\ref{fig:gamma} illustrates the transmit power achieved by different schemes versus the NLOS user's rate constraint $\Gamma_{\rm N}$. The transmit power increases with $\Gamma_{\rm N}$ for all schemes, with ``Near LU" consistently exhibiting the worst performance due to the severe path loss of the cascaded NLOS link. When $\Gamma_{\rm N}$ is low, the ``Midpoint" scheme outperforms ``Near RIS" by better balancing the channel conditions of the two users. As $\Gamma_{\rm N}$ increases, ``Near RIS" becomes more effective by improving the weaker NLOS link to satisfy the high rate constraint. The results highlight the importance of adapting PA positions to different user rate requirements. By further optimizing the PA positions and RIS reflection coefficients, the proposed AO method better exploits the spatial DoFs to enhance the effective channels and mitigate inter-user interference, achieving the lowest transmit power with a 1.28 dB average reduction compared with the ``1-D search" scheme. It can also be observed that as the NLOS user's rate requirement increases, the optimized transmit power approaches that of the ``Near RIS scheme", indicating that the severe path loss of the NLOS link becomes a dominant factor in PA position optimization.

\begin{figure}[t]
	\vspace{-1.5mm}
	\centering{\includegraphics[width=0.8\columnwidth]{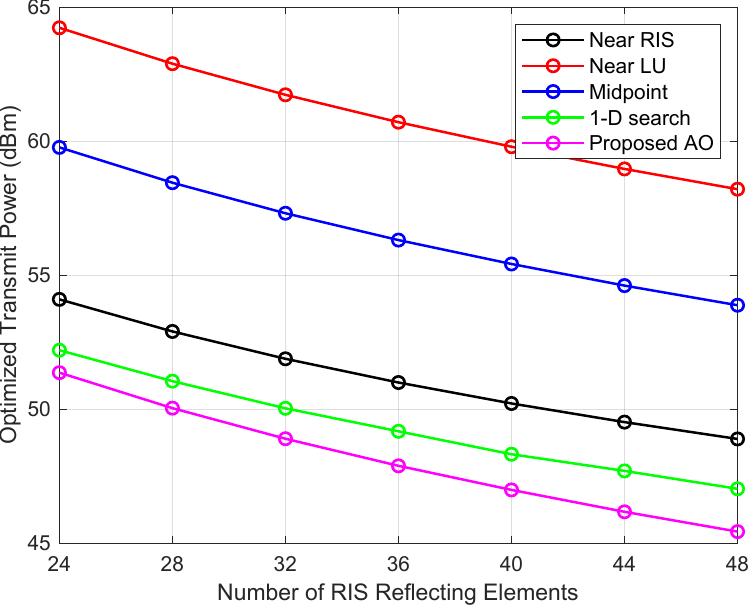}}
	\caption{The optimized transmit power versus the number of RIS reflecting elements with $\Gamma_{\rm L}=8$ bps/Hz and $\Gamma_{\rm N}=4$ bps/Hz.}
	\vspace{-1.5mm}
	\label{fig:L}
\end{figure}
The transmit power achieved by different schemes versus the number of RIS reflecting elements $L$ is shown in Fig.~\ref{fig:L}. As $L$ increases, the additional RIS phase-shift DoFs enhance the power minimization performance. However, heuristic schemes that simply place PAs near the LOS user or at the midpoint fail to fully exploit the available spatial DoFs, resulting in inferior performance compared with the ``Near RIS" scheme, which focuses on reducing the path loss of the NLOS link. The proposed AO method consistently achieves the lowest transmit power by jointly optimizing the PA positions and RIS reflection coefficients. Moreover, its advantage over the ``1-D search" scheme becomes more pronounced as $L$ increases, achieving power reductions of 0.83 dB and 1.60 dB at $L=24$ and $L=48$, respectively.

\begin{figure}[t]
	\centering{\includegraphics[width=0.8\columnwidth]{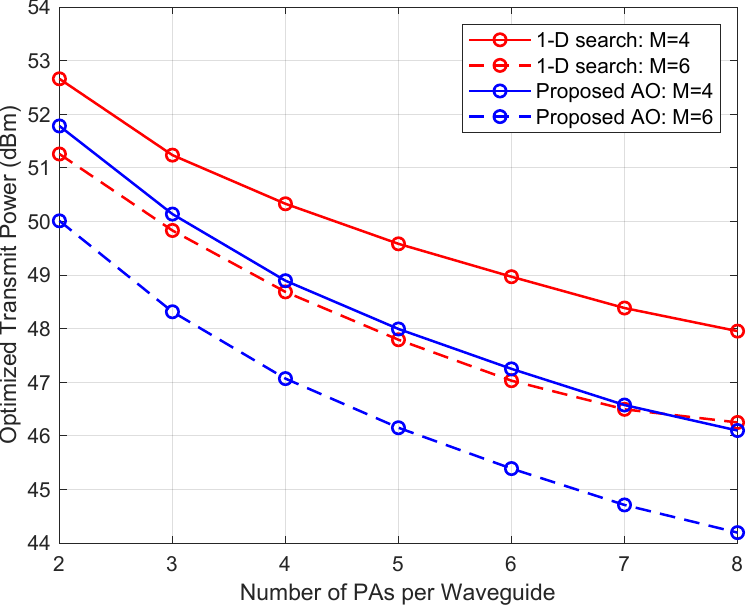}}
	\caption{The optimized transmit power versus the number of PAs per waveguide with $\Gamma_{\rm L}=8$ bps/Hz and $\Gamma_{\rm N}=4$ bps/Hz.}
	\vspace{-1.5mm}
	\label{fig:MN}
\end{figure}

Figure~\ref{fig:MN} illustrates the transmit power achieved by the two joint optimization schemes for different numbers of waveguides $M$ and PAs per waveguide $N$. As $M$ and $N$ increase, the additional spatial DoFs lead to lower transmit power. The proposed AO method consistently outperforms the 1-D search scheme, demonstrating its effectiveness in exploiting the additional spatial DoFs. This is because the element-wise search optimizes one PA position at a time, limiting its ability to jointly improve the overall channel structure. In contrast, the proposed AO method simultaneously updates all PA positions along a joint gradient-based direction, thereby better exploiting the additional DoFs to reduce transmit power.

\begin{table}[t]
	\centering
	\caption{Average Execution Time of Different Schemes (s)}
	\label{tab:time}
	\renewcommand{\arraystretch}{1.15}
	\setlength{\tabcolsep}{7pt}
	\begin{tabular}{c c c c}
		\hline
		$L$ & $(M,N)$ & 1-D Search & Proposed AO \\
		\hline
		24 & (4,4) & 0.94 & 0.12 \\
		48 & (4,4) & 1.83 & 0.37 \\
		48 & (6,6) & 7.97 & 0.98 \\
		\hline
	\end{tabular}
	\vspace{-1.5mm}
\end{table}

Table~\ref{tab:time} compares the average execution time of the two joint optimization schemes under different system configurations. The execution time of the ``1-D search" scheme increases substantially with the number of RIS elements, waveguides, or PAs, due to the increased computational cost of repeated element-wise searches. In contrast, the proposed AO method simultaneously updates all PA positions using joint gradient information, substantially reducing the growth in execution time and making it more suitable for practical implementation.

\section{Conclusion}

In this paper, we have investigated an RIS-assisted PASS system in the presence of NLOS users, where the RIS extends the coverage of PASS to simultaneously serve LOS and NLOS users. We adopted ZF beamforming to satisfy the communication rate constraints and proposed an AO framework to jointly optimize the RIS reflection coefficients and PA activation positions for transmit power minimization. Simulation results demonstrated that joint optimization significantly outperforms heuristic PA deployment schemes. Moreover, the proposed AO method achieves lower transmit power with substantially reduced execution time compared with the existing element-wise 1-D search scheme. Future work will investigate the joint optimization with more advanced beamforming designs and extend the proposed framework to multiuser scenarios with multiple LOS and NLOS users.

\footnotesize
\bibliographystyle{IEEEtran}
\bibliography{Refs}{}

\begin{thebibliography}{10}
\providecommand{\url}[1]{#1}
\csname url@samestyle\endcsname
\providecommand{\newblock}{\relax}
\providecommand{\bibinfo}[2]{#2}
\providecommand{\BIBentrySTDinterwordspacing}{\spaceskip=0pt\relax}
\providecommand{\BIBentryALTinterwordstretchfactor}{4}
\providecommand{\BIBentryALTinterwordspacing}{\spaceskip=\fontdimen2\font plus
\BIBentryALTinterwordstretchfactor\fontdimen3\font minus
  \fontdimen4\font\relax}
\providecommand{\BIBforeignlanguage}[2]{{%
\expandafter\ifx\csname l@#1\endcsname\relax
\typeout{** WARNING: IEEEtran.bst: No hyphenation pattern has been}%
\typeout{** loaded for the language `#1'. Using the pattern for}%
\typeout{** the default language instead.}%
\else
\language=\csname l@#1\endcsname
\fi
#2}}
\providecommand{\BIBdecl}{\relax}
\BIBdecl

\bibitem{ra_sur}
W.~Ma \emph{et~al.}, ``{A survey on reconfigurable and movable antennas for
  wireless communications and sensing},'' \emph{IEEE Commun. Surv. Tutor.},
  vol.~28, pp. 4842--4882, 2026.

\bibitem{ris_sur}
Y.~Liu \emph{et~al.}, ``{Reconfigurable intelligent surfaces: Principles and
  opportunitie}s,'' \emph{IEEE Commun. Surv. Tutor.}, vol.~23, no.~3, pp.
  1546--1577, 2021.

\bibitem{pa_sur}
Y.~Lu \emph{et~al.}, ``{A Survey on pinching-antenna systems: Optimization
  techniques, intelligent designs, and multifunctional applications},''
  \emph{IEEE Commun. Surv. Tutor.}, vol.~28, pp. 6592--6627, 2026.

\bibitem{ris1}
Q.~Wu and R.~Zhang, ``{Intelligent reflecting surface enhanced wireless network
  via joint active and passive beamforming},'' \emph{IEEE Trans. Wirel.
  Commun.}, vol.~18, no.~11, pp. 5394--5409, 2019.

\bibitem{pa1}
Y.~Liu \emph{et~al.}, ``{Pinching-antenna systems: Architecture designs,
  opportunities, and outlook},'' \emph{IEEE Commun. Mag.}, vol.~64, no.~1, pp.
  190--196, 2026.

\bibitem{pa2}
Z.~Ding and H.~V. Poor, ``Analytical optimization for antenna placement in
  pinching-antenna systems,'' \emph{IEEE Trans. Wirel. Commun.}, vol.~25, pp.
  21\,943--21\,957, 2026.

\bibitem{pa3}
W.~Huang, Y.~Lin, K.~C. Teh, O.~A. Dobre, and Z.~Ding, ``{Power minimization
  for pinching-antenna systems: Flexible hybrid NOMA perspective},'' \emph{IEEE
  Trans. Commun.}, vol.~74, pp. 13\,050--13\,064, 2026.

\bibitem{pa4}
J.~Zhao \emph{et~al.}, ``{Pinching-antenna systems-enabled multi-user
  communications: Transmission structures and beamforming optimization},''
  \emph{IEEE Trans. Commun.}, vol.~74, pp. 2316--2330, 2026.

\bibitem{pa5}
M.~Sun, C.~Ouyang, S.~Wu, and Y.~Liu, ``Multiuser beamforming for
  pinching-antenna systems: An element-wise optimization framework,''
  \emph{IEEE Trans. Wirel. Commun.}, vol.~25, pp. 6538--6552, 2026.

\bibitem{pa_ris1}
M.~Samy \emph{et~al.}, ``Pinching antenna systems versus reconfigurable
  intelligent surfaces in mmwave,'' \emph{IEEE Wirel. Commun. Lett.}, vol.~14,
  no.~12, pp. 4022--4026, 2025.

\bibitem{pa_ris2}
X.~Fu, L.~Chen, L.~Zheng, and L.~Gao, ``{Rate maximization for RIS-assisted
  downlink pinching-antenna systems},'' \emph{IEEE Wirel. Commun. Lett.},
  vol.~15, pp. 4484--4488, 2026.

\bibitem{pa_ris3}
C.~He, Y.~Lu, Y.~Xu, C.-Y. Chi, and A.~Nallanathan, ``{RIS-assisted downlink
  pinching-antenna systems: GNN-enabled optimization approaches},'' \emph{arXiv
  preprint arXiv:2511.20305}, 2025.

\bibitem{precode}
A.~Wiesel, Y.~Eldar, and S.~Shamai, ``{Linear precoding via conic optimization
  for fixed MIMO receivers},'' \emph{IEEE Trans. Signal Process.}, vol.~54,
  no.~1, pp. 161--176, 2006.

\bibitem{rcg1}
R.~Liu, M.~Li, Q.~Liu, and A.~L. Swindlehurst, ``{Joint symbol-level precoding
  and reflecting designs for IRS-enhanced MU-MISO systems},'' \emph{IEEE Trans.
  Wirel. Commun.}, vol.~20, no.~2, pp. 798--811, 2020.

\bibitem{rcg2}
Y.~Geng, T.~Hiang~Cheng, K.~Zhong, and K.~Chan~Teh, ``{Unified manifold
  optimization for double-IRS-aided MIMO communication},'' \emph{IEEE Commun.
  Lett.}, vol.~28, no.~7, pp. 1713--1717, 2024.

\bibitem{HS}
H.~Sato, ``{Riemannian conjugate gradient methods: General framework and
  specific algorithms with convergence analyses},'' \emph{SIAM J. Optim.},
  vol.~32, no.~4, pp. 2690--2717, 2022.

\bibitem{BB}
\BIBentryALTinterwordspacing
E.~G. Birgin, J.~M. Martínez, and M.~Raydan, ``{Spectral projected gradient
  methods: Review and perspectives},'' \emph{J. Stat. Softw.}, vol.~60, no.~3,
  pp. 1--21, 2014. [Online]. Available:
  \url{https://www.jstatsoft.org/index.php/jss/article/view/v060i03}
\BIBentrySTDinterwordspacing

\bibitem{PAVA}
A.~Henzi, A.~M{\"o}sching, and L.~D{\"u}mbgen, ``Accelerating the
  pool-adjacent-violators algorithm for isotonic distributional regression,''
  \emph{Methodol. Comput. Appl. Probab.}, vol.~24, no.~4, pp. 2633--2645, 2022.

\end{thebibliography}

\end{document}